\documentclass[a4paper,fleqn]{elsart}
\usepackage{amsmath}

\begin{document}

\begin{frontmatter}
\title{On the quantum measurements from a reconsidered perspective}
\author{S.Dumitru}\footnote{Electronic mail: s.dumitru@unitbv.ro}
\address{Department of Physics, ``Transilvania'' University, B-dul Eroilor 29, 
R-2200 Bra\c{s}ov, Romania}
\date{\today}
\maketitle
\begin{abstract}
For theoretical approach of quantum measurements it is proposed a set of reconsidered
 conjectures. The proposed approach implies linear functional transformations for
probability density and current but preserves the expressions for operators of observables.
The measuring uncertainties appear as changes in the probabilistic estimators of
 observables.
\end{abstract} 
\end{frontmatter}

\section{Introduction}

The question of a theory of measurements is present \cite{1,2,3} in
 many debates about
quantum mechanics (QM) but \cite{4} it did not exist prior to QM. The respective question 
arose from the discussions on the traditional interpretation of uncertainty 
relations (TIUR)) and it  has generated a large diversity of viewpoints regarding 
its importance 
and approach. The respective diversity inserts even some extreme opinions such are:
\begin{itemize}
\item \textbf{(i)} the description of quantum measurements (QMS)is \emph {"probably 
the most important part of the theory"} \cite{1}.
\item \textbf{(ii)} :\emph{"the word (\emph{'measurement'})has had such a damaging effect on 
the discussions
that ...  it should banned altogether in quantum mechanics"} \cite{5}.
\end{itemize}

As a notable aspect today one finds that many of the existing approaches of QMS (including
some  of the most recent ones) are TIUR-connected, because they are founded on traditional
conjectures inspired  someway from TIUR. In their essence the respective approaches, as well as the
TIUR itself, are based on the idea that the Robertson Schr\"{o}dinger uncertainty 
relation (RSUR) is a capital physical formula with a straightforward significance for QMS.
But a minute re-examination of the facts shows that TIUR is nothing but an unjustified 
doctrine respectively that RSUR is a simple fluctuation formula without any significance
for QMS (see the investigations progressively developed in our works \cite{6,7,8,9,10}).

In the mentioned circumstances it becomes of real and actual interest for the theory of QMS
to search new approaches which are disconnected of TIUR doctrine. Such an approach is the 
aim of the present paper. For our aim in the next section we present the deficiencies of 
the traditional 
conjectures regarding the QMS approaches. In section 3 we argue for a set of reconsidered 
conjectures regarding the QMS theory. Subsequently in section 4 we propose a new approach 
of QMS. The proposal is inspired from a view \cite{11,12} about the measurement 
of classical (non-quantum) random observables. The proposed approach is detailed through 
a simple exemplification in the section 5. Some ending conclusions are given in section 6.
\section{Traditional conjectures and their deficiencies}

In its essence TIUR is reducible \cite{10} to a set of main assertions for which the 
reference element is the RSUR 
\begin{equation}\label{eq:1}
\Delta \,A \cdot\, \Delta \,B\;\ge \; \frac{1}{2}\;|\,\langle[\hat A,\hat B ]\,\rangle|
\end{equation}
taken as a capital formula of physics. According to the mentioned assertions $\Delta A$
and $\Delta B$ are considered as uncertainties in simultaneous measurements of observables
$A$ and $B$. The commutator $[\hat A\,,\hat B ]$ is considered as a distinction sign 
between the cases of compatible and non-compatible observables (when $[\hat A\,,\hat B 
] = 0$ , respectively $[\hat A\,,\hat B ] \neq 0$). The mentioned considerations are 
associated with the idea of non-null and unavoidable perturbations attributed  to QMS.
The asserted characteristics of RSUR \eqref{eq:1} and the associated perturbations are
considered to be specific only for QMS and without analogues in classical physics
(for more details see \cite{10}).

Inspired from the alluded  considerations the TIUR-connected approaches of QMS imply 
someway one or more of the following traditional conjectures (\textbf{TC}):
\begin{itemize}
\item \textbf{TC.1}: The descriptions of QMS must be developed as extensions/completions 
of TIUR. Subsequently the respective descriptions have to contain distinctive elements for 
the pairs of compatible respectively non-compatible observables.
\item \textbf{TC.2}: the quantum and classical measurement must be approached with 
essentially distinct views because of the exclusive character of RSUR \eqref {eq:1} and 
of the QMS perturbations.
\item \textbf{TC.3}: The descriptions of QMS have to take into account the jumps caused 
by the mentioned perturbations on the states of the measured systems.
\textbf{TC.4}: A QMS consists in a single trial (detection act) that gives as result 
an unique value for the measured observable. Consequently its description must be 
associated somehow with a collapse/reduction of the (random content of) corresponding
wave function.
\end{itemize}
The mentioned re-examination of the facts shows \cite{10} that in reality RSUR \eqref{eq:1}
 is not a capital formula of physics. Therefore one can conclude that it is 
 unreasonable to subordinate the QMS approaches to TIUR. But such a conclusion clearly 
 invalidates the conjecture \textbf{TC.1}. On the other hand through the same 
 re-examination one finds that the RSUR \eqref {eq:1} belongs to a general family of 
 fluctuations formulas from both quantum and classical (non-quantum) physics. 
 Then it results that there are not real motives for a fundamental distinction between quantum and classical 
situations. Evidently that such a result leaves without any base the conjecture
\textbf{TC.2}.

The conjecture \textbf{TC.3} is also TIUR-connected. Firstly it was said that the 
uncertainties put forward by TIUR are due to the interactions between measuring devices 
and measured systems. Secondly it was added that the respective interactions cause jumps 
in the states of measured systems. Then it was promoted the idea that, in contrast 
with the classical situations, in QMS the mentioned uncertainties, interactions and jumps
have an unavoidable character. Subsequently it was promoted the idea that the alluded 
jumps must be taken into account in the descriptions of QMS. But, in spite of its genesis,
the respective idea is proved to be incorrect by the following natural and indubitable 
remark \cite{13}: \emph{"it seems essential to the notion of measurement that it answers a 
question about the given situation existing before the measurement. Whether the measurement
leaves the measured system unchanged or brings about a new and different state of that 
system is a second and independent question".} So we have to report an indubitable 
deficiency of \textbf{TC.3}.

The conjecture \textbf{TC.4} is contradicted by the well-known character of random 
variables manifested by the quantum observables. Indeed, mathematically 
\mbox{\cite{14,15,16}}, 
in a given situation a random variable is characterized not by a unique value
but by a spectrum of values. Consequently for a random observable, in a given state 
of the considered physical system a single trial has no significance.  For a true 
estimation of such an observable a significant measurement must consist in a statistical 
sampling composed by an (adequately) large number of individual trials. The mentioned 
features of  random observables are correctly taken into account in the frame of
the classical physics (e.g. in the study of the fluctuations), where to the spectrum of a
 random observable is attached the corresponding probability distribution.
  In the same frame a measurement of a random observable is not associated with 
  a collapse/reduction of the respective probability distribution. The association can 
  be done \cite{11,12} through a functional transformation of the mentioned 
distribution. Then, because the quantum observables have random characteristics, 
it results directly that a QMS must be regarded as a statistical sampling (in the above
noted sense). Consequently there are no reasons to represent (describe)a QMS by means 
of collapse/reduction of wave function. So we can conclude that the conjecture
\textbf{TC.4 } is incorrect.

The above presented facts reveal visible deficiencies of traditional conjectures 
\textbf{TC.1-4}. The respective deficiencies  have an unsurmountable  character because 
they cannot be remedied (or avoided) by valid arguments derivable from the TIUR doctrine. 
Then it results that TIUR-connected approaches of QMS are attempts with wrong  grounds. 
Due to the persistence of such attempts in the nowadays publications it result that , at least 
partially, the problem of QMS description is a still open question which requires new 
approaches founded on reconsidered conjectures and disconnected of TIUR doctrine. Such an
approach we try to present in the next sections.

\section{Arguments for reconsidered conjectures}
A natural theory of measurements must contain elements (conjectures, reasonings) which 
are in an adequate correspondence with the main characteristics of the real measuring 
experiments. Then, for our trial, it is of first interest, to note the respective 
characteristics for QMS regarded in the general context of experimental physics.

In classical physics the belief \cite{4}: \emph{"in the objective existence of material 
systems ... which possesses properties independently of measurements"} is accepted as
 an axiom. The respective axiom implies the idea that a measurement aims to give 
information about the pre-existent state of the investigated system. We opine that the 
mentioned axiom and idea must be adopted also in connection with the quantum systems.
Our opinion is encouraged by the following indubitable remarks expressed with regard 
to QMS :
\begin{itemize}
\item \textbf{(i)} \emph{"When it is said that something is \emph{'measured'} it is difficult  
not to think of the result 
are referring to some preexisting property of the object in question"}\cite{5}, \newline
respectively.
\item \textbf{(ii)} a measurement \emph{"answer a question about the given situation existing 
before the measurement"} \cite{13}.
\end{itemize}
In the quantum situations it is also significant the fact that, from a mathematical 
viewpoint, the observables are random variables and consequently 
their measurements must consist in statistical samplings. The essential aspects of the
respective fact were pointed out in the previous section in discussions about the 
conjecture  \textbf{TC.4}. Such aspects have to be integrated in both experimental and 
theoretical approaches of QMS. 

By regarding things as above an experimental approach of QMS can be described as follows:
Let be a QMS consisting  of $N$ single trials destined to measure concomitantly two 
observables $A$ and $B$. All trials operate on the same pre-existent state of the 
considered system (or of its identical replicas). Corresponding to the respective state
for $A$ and $B$ the $i-th$ trial gives the values $\alpha_i$ respectively $\beta_i$
($i = 1, 2, ..., N$). Then, according to the mathematical statistics \cite{14,15,16},
the observables $A$ and $B$ are evaluated through the factual (\emph{fac}) estimators: mean 
values $\langle A \rangle _{fac} = \langle \alpha \rangle$ , $\langle B 
\rangle _{fac}  = \langle \beta \rangle$ , correlation 
$\mathcal{C}_{fac}(A,B)$ and standard deviations $\Delta_{fac}A$, $\Delta_{fac}B$. With the notations
$\delta \alpha _{i}=\alpha _{i}-\langle \alpha \rangle $ and $\delta \beta _{i}
=\beta _{i}-\langle \beta \rangle$ the mentioned estimators are defined by 
the relations :
\begin{equation}\label{eq:2}
\langle A \rangle _{fac}  = \langle\alpha\rangle=
\frac{1}{N}\,\sum\limits_{i = 1}^N {\alpha _i } 
\quad ,\quad   \langle B \rangle _{fac}  = \langle\beta\rangle =
\frac{1}{N}\,\sum\limits_{i = 1}^N {\beta _i } 
\end{equation}
\begin{equation}\label{eq:3}
\mathcal{C}_{fac} (A,B) = \frac{1}{N}\,\sum\limits_{i = 1}^N {\delta \alpha _i \, \cdot \,} 
\delta \beta _i 
\end{equation}
\begin{equation}\label{eq:4}
(\Delta _{fac} A)^2  = \frac{1}{N}\,\sum\limits_{i = 1}^N 
{(\delta \alpha _i )^2 } \quad ,\quad \quad (\Delta _{fac} \,B)^2  = 
\frac{1}{N} \sum\limits_{i = 1}^N {(\delta \beta _i )^2 } \quad 
\end{equation}
Let us remark that in mathematical statistics when an observable $A$ is a deterministic  
(non-random) variable the quantity $\langle A\rangle_{fac}$ estimate 
an intrinsic characteristic 
of the respective observable while
$\Delta_{fac} A$ describe the error (uncertainty) with which it is 
evaluated $\langle A\rangle_{fac}$.
But in the case when $A$ is a random variable both $\langle A\rangle_{fac}$ 
and $\Delta_{fac} A$ evaluate
intrinsic characteristics of $A$. In such a case the errors are specific at the same 
time for both $\langle A \rangle_{fac}$ and $\Delta_{fac}\,A$ . The respective 
errors can be described by
means of some uncertainty indicators $\varepsilon [\langle A \rangle_{fac}]$ 
and $\varepsilon [\Delta_{fac} A]$
defined as
\begin{equation}\label{eq:5}
\varepsilon [\langle A \rangle_{fac}] = 
\{ V[\, \langle A \rangle_{fac} ]\} ^{1/2} \quad ,
\quad \quad \varepsilon [\Delta _{fac} A] =
\{ V[\Delta_{fac} A]^2 \} ^{1/4} 
\end{equation}
where $V \,[\eta]$ denote the variance of $\eta$ defined in terms of $\alpha_i$ and
$\beta_i$ by means of usual formulas from  mathematical statistics \cite{14,15}.
We do not reproduce the respective formulas because here they are not of direct 
necessity.

Now let us observe that from the total number $N$ of values $\alpha _1, \alpha _2, ...,
\alpha _N$ only a restricted number $n$ $(n \,< \,N)$ of them are distinct. The respective 
distinct values , denoted by $a_1, a_2, ... ,a_n$ constitute the spectrum of $A$.
Such a value $a_i$ appears in $N_i$ trials and , consequently it is associated with the 
selection frequency $\nu _j  = \frac{{N_j }}{N}$ (evidently 
$\sum\limits_{j = 1}^n {\nu _j }  = 1$). So for $\langle A \rangle_{fac}$ 
and $\Delta_{fac} A$  one 
obtains the expressions : 
\begin{equation}\label{eq:6}
 \langle A \rangle _{fac} = \sum\limits_{j = 1}^n {\nu _j }  \cdot a_j \quad ,
 \quad \quad \Delta _{fac} A = [\sum\limits_{j = 1}^n {\nu _j } (a_j  - 
  \langle A\rangle  _{fac} )^2 ]^{1/2} 
\end{equation}
These expressions allow to point out the idea of spectrum preservation often assumed in 
experimental practice. According to the respective idea in measurements of a random 
quantity $A$ the changes of experimental performances aim ameliorative transformations
for the frequencies $\nu _j$ but assume the preservation of the values $a _j$ which
define the spectrum of $A$.\\
Based on the above considerations we think that a natural approach of QMS can be founded
on the following reconsidered conjectures (\textbf{RC}):
\begin{itemize}
\item \textbf{RC.1}: The purpose of a QMS is to give information about the pre-existent
state of the investigated system.
\item \textbf{RC.2}: Because QMS view the systems studied in QM, their theoretical 
descriptions must be done in terms of quantum operators and wave functions.
\item \textbf{ RC.3}: A QMS consists in a statistical sampling which preserves
the spectra of observables but can modify the corresponding probabilities. That is why 
theoretically a QMS must be described as an operation which preserves the mathematical
expressions of operators but which is associated with functional transformations for the 
wave functions (or for related probabilistic quantities). So the description of QMS is 
dissociated of any idea  about the collapse (reduction) of wave function.
\item \textbf{RC.4}: Since the simple usual QM refers only to the intrinsic properties of
quantum systems the transformations mentioned in \textbf{RC.3} must contain some extra-QM 
elements regarding the measuring devices and procedures. Then the description of QMS 
appears not as a part of QM theory but as a distinct and independent task comparatively
with the objectives of usual QM.
\end{itemize}

\section{An  approach from the reconsidered perspective}
Now let us develop a theoretical model for description of a QMS regarded as as an 
operation with the characteristics announced in \textbf{RC.1-3}.

To the respective characteristics we add firstly the observation that the wave functions 
incorporate information (of probabilistic nature) about the measured system. That is why 
a QMS can be regarded as a process of information transmission, from the respective 
system to the recorder of the measuring device. The input (\emph{in}) information,
described by a wave function $\psi_{in}$, regards the measured system considered as information 
source. The output (\emph{out}) information, described by a wave 
function $\psi_{out}$, refers 
to the data received on the recorder taken as information receiver. So the measuring 
device plays the role of a channel for information transmission. Then the errors 
(uncertainties)induced by measurement appears as alteration of the transmitted 
information.

As a matter of fact a QMS description of the mentioned kind can be depicted as follows.
Let be a spinless microparticle (quantum system) whose own characteristics of orbital type are
described by the intrinsic (\emph{in}) wave function $\psi _{in}(\vec{r})$, regarded as 
solution of the corresponding Schr\"{o}dinger equation. The observables of interest 
(such are coordinates, the components of linear and angular momenta or energy) will 
be designed by $A _k \, (k = 1, 2, ..., s)$.They are described by the corresponding 
usual QM operators $\hat A _k$ regarded as generalized random variables. In spirit of
the discussions from section 3 we suggest that within the theoretical description the
QMS operation must leave unchanged the spectra (and consequently the mathematical 
expressions) of operators $\hat A _k$. The same operation has to transform the quantum 
probabilities from \emph{in}-reading into \emph{out}-reading. The mentioned probabilities 
are associated with the densities $\varrho _z$ and currents $\vec J _z$ 
($z = in, out$)defined by:
\begin{equation}\label{eq:7}
\rho _z  = \left| \psi _z  \right| ^{\,2} \quad ,\quad \quad 
\quad \vec{J}_z  = \frac{\hbar }{m}\left| \psi _z \right|^{2} 
\cdot \nabla \phi _z 
\end{equation}
Here $\left| \psi _z  \right|$ and $\phi _z$ denote the modulus respectively the argument 
of $\psi _z$ (i.e. $\psi _{z} = \left| \psi _z \right|\cdot exp (i \phi _z)$) and $m$ 
represents the particle mass. 

The alluded association is connected with the facts that \cite {17} the set $\varrho _z$ and  
$\vec J _z$ \emph{ "encodes the probability 
distributions of quantum mechanics"} and it \emph{"is in principle measurable by virtue 
of its effects on other systems"}. To be added here the possibility \cite {18} for taking 
in QM as primary entity the set  $\varrho _{in}$ - $\vec J _{in}$  but not the wave 
function $ \psi _{in}$ (i.e. the possibility to start the QM considerations with the 
continuity equation for  $\varrho _{in}$ - $\vec J _{in}$ and subsequently to derive 
the Schr\"{o}dinger equation for $ \psi _{in}$).

The above noted  observations suggest that the transformations 
$\psi _{in} \rightarrow \psi _{out} $ to be described in terms of $\varrho _z$ and 
 $\vec J _z$ ($z = in, out$). But  $\rho _z$ and $\vec{J}_z$ refer to the position 
respectively motion kinds of probabilities. Experimentally the two kinds  are regarded as
measurable by distinct devices and procedures. Then the aimed description of QMS has 
to combine the distinct functional transformation of $\rho _{in}$ in  $\rho _{out}$
respectively of $\vec{J}_{in}$ in  $\vec{J}_{out}$.

Similarly with the classical situations \cite{11,12}, for 
 which are implied measuring devices with linear and stationary characteristics, the alluded 
 functional transformations can be written as :
\begin{equation}\label{eq:8}
\rho _{out} (\vec r)= \int {\Gamma (\vec r,\vec {r}\,')} \rho _{in} 
(\vec {r}\,')\,d^{3} \vec{r}\,'
\end{equation}
\begin{equation}\label{eq:9}
J_{out;\, k} (\vec r) = \sum\limits_{l = 1}^3 \int {\Lambda _{kl} }  
(\vec r , \vec {r}\,') 
J_{in;\,l} (\vec {r}\,')\, d^{\,3} \vec {r}\,' 
\end{equation}
( $J_{z;\,k}$ with $z = in, out$ and $k = 1,2,3 = X,Y,Z$ denote the 
cartesian components of $\textbf{J}_{z})$. Due to the probabilistic  
correspondence between  $\rho _{in}$ 
and
$\rho _{out}$  
respectively between $\textbf{J} _{in}$ and $\textbf{J} _{out}$ the
kernels $\Gamma (\vec r, \vec{r}\,')$ and $\Lambda _{kl} (\vec r, \vec{r}\,')$ satisfy the
conditions :
\begin{equation}\label{eq:10}
\int {\Gamma (\vec r,\vec {r}\,'})\,d^3 \vec r = \int {\Gamma (\vec r,
\vec {r}\,'})\,d^3 \vec {r}\,' = 1 
\end{equation}
\begin{equation}\label{eq:11}
\sum\limits_{k = 1}^3 {\int {\Lambda _{k\,l} (\vec r,\vec {r}\,' } } )\, d^3 \vec r 
= \sum\limits_{l = 1}^3 {\int {\Lambda _{kl} (\vec r,\vec {r}\,'}} 
)\,d^3 \vec {r}\,' =1
\end{equation}
The kernels $\Gamma$ and $\Lambda _{k l}$ describe the transformations induced by QMS
in the information regarding the measured system (microparticle). Therefore they must
incorporate some extra-QM elements regarding the characteristics of the measuring devices 
and procedures. The respective elements do not belong to the usual QM description for 
the intrinsic properties of the measured system. 

The above considerations facilitate an evaluation of the effects induced by QMS on the 
probabilistic estimators of orbital observables specific to the measured system. Such
observables are described by the operators $\hat A _k$ whose expressions depend on
$\vec r$ and $\nabla$. According with \textbf{RC.3} the mentioned expressions are 
supposed to remain invariant under the transformations which describe QMS. So one can say 
that in the situations associated with the wave functions $\psi _{z}$ ($z = in, out$)
two observables $A$ and $B$ are described by the following probabilistic estimators :
mean values $\langle A \rangle _{z}$ and $\langle B \rangle _{z}$, 
correlation $C _{z}(A,B)$ and standard deviations
$\Delta _{z}A$ and $\Delta _{z}B$. With the usual notation 
$(f,g) = \int f^* g\,d^3 \vec r$ for the product of two functions $f$ and $g$,
the mentioned estimators are defined by the relations:
\begin{equation}\label{eq:12}
 \langle A \rangle _z = (\psi _z ,\hat A\psi _z )\quad ,\quad \quad 
  \langle B \rangle _z  = (\psi _z ,\hat B\psi _z )
\end{equation}
\begin{equation}\label{eq:13}
\mathcal{C}_z (A,B) = (\delta _z \hat A \psi _{z} ,\delta_z \hat B \psi_{z})
,\quad \delta _z \hat A = \hat A -  \langle A \rangle _z 
\end{equation}
\begin{equation}\label{eq:14}
\Delta _z X = \sqrt {\mathcal{C}_z (X,X)} \quad ,\quad \quad X = A, B
\end{equation}
Note that the estimators \eqref{eq:12} - \eqref{eq:14} can be calculated by utilization
of the basic probability elements $\varrho _{z}$ and $\textbf{J} _{z}$. So if $\hat A$ does not depend
on $\nabla$ (i.e. $\hat A = A(\vec r)$) in evaluating the scalar products from
\eqref{eq:12} - \eqref{eq:14}  one can use the evident equality $\psi_z^*  
\hat A \psi _{z}  = A(\vec {r})\rho_z $. When $\hat A$ depends on $\nabla$
(i.e  $\hat A = A(\nabla)$) in the same products can be appealed the substitutions
\begin{equation}\label{eq:15}
\psi_{z}^*  \nabla \psi_z  = \frac{1}{2} \nabla \rho_z  + 
\frac{im}{\hbar } \vec J_z 
\end{equation}
\begin{equation}\label{eq:16}
\psi_z^*  \nabla^2 \psi _z = \rho _z^{1/2} \nabla^2 
\rho _z^{1/2}  + \frac{im}{\hbar }\nabla \vec J_z  - \frac{m^2 }
{\hbar ^2}\frac{\vec{J_z}^2 }{\rho_z}
\end{equation}
For the evaluation of the estimators \eqref{eq:12}, \eqref{eq:13} and \eqref{eq:14} the
above mentioned utilization seems to allow the avoidance the implications of \cite {17}
\emph{"a possible nonuniqueness of current 
\emph{(i.e. of the set $\varrho_z$ and $\vec J_z)$}"}.
Within the above approach, for two observables $A$ and $B$, the measuring errors can be
evaluated through the following predicted (\emph{prd})  uncertainty indicators: 
\begin{equation}\label{eq:17}
\varepsilon _{prd}\left( {\left\langle A \right\rangle } \right) = \left| 
{\left\langle A \right\rangle _{out} - \left\langle 
A \right\rangle _{in} } \right|
\end{equation}
\begin{equation}\label{eq:18}
\varepsilon_{prd} \left( {\mathcal{C}\left( {A,B} \right)} \right) = \left| 
\, \mathcal{C}_{out} \left( {A,B} \right) - \mathcal{C}_{in} \left( {A,B}
 \right) \right|
\end{equation}
\begin{equation}\label{eq:19}
\varepsilon_{prd} \left( {\Delta A} \right) = \left| {\Delta _{out} A - 
\Delta _{in} A} \right|
\end{equation}
These indicators incorporate errors regarding the whole spectrum of the observables
$A$ and $B$ considered as random variables.

Now note that the $z = out$ version of probabilistic indicators 
\eqref{eq:12}-\eqref{eq:14} as well as uncertainties indicators 
\eqref{eq:15}-\eqref{eq:17}  have a theoretical
significance. The adequacy of such estimators and indicators must be tested by comparing
them with their factual (\emph{fac}) correspondents \eqref {eq:2}-\eqref {eq:4}
respectively \eqref {eq:5}.
If the test is affirmative  both theoretical descriptions, of QM intrinsic properties
 of system and of QMS, can be accepted as adequate. But if  test gives an invalidation, 
 at least one of the mentioned description must be regarded as inadequate. 
 
 \section {A simple exemplification} 
  For an exemplification of the above considerations let us refer to a microparticle in a
  one-dimensional motion along the x-axis. We take $\psi _{in} \left( x \right) = 
  \left| {\psi _{in} \left( x \right)} \right| \cdot \exp \left\{ {i\phi _{in} } 
  \right\}$ with
  \begin{equation}\label{eq:20}
\left| {\psi _{in} \left( x \right)} \right| = \left( {\sigma \sqrt {2\pi } }
 \right)^{- 1/4} \cdot \exp \,\left\{ { - \frac{{(x - x_0 )^2 }}{{4\sigma ^2 }}}
  \right\}\quad ,\quad \phi _{in} \left( x \right)=k \cdot x
\end{equation}
Correspondingly we have
  \begin{equation}\label{eq:21}
\rho _{in} \left( x \right) = \left| {\psi _{in} \left( x \right)} \right|^2 ,
\quad J_{in} \left( x \right) = \frac{{\hbar k}}{m}\left| {\psi _{in} 
\left( x \right)} \right|^2 
\end{equation}
So the intrinsic characteristics of the microparticle are described by the parameters 
$x_0$, $\sigma$ and $k$.

If the errors induced by QMS are small in \eqref{eq:8}-\eqref{eq:9} we can operate with 
the kernels of Gaussian forms given by:
  \begin{equation}\label{eq:22}
 \Gamma \left( {x,x'} \right) = \left( {\gamma \sqrt {2\pi } } \right)^{-1} 
 \cdot \exp \left\{ { - \frac{{\left( {x - x'} \right)^2 }}{{2\gamma ^2 }}}
  \right\}
 \end{equation}
  \begin{equation}\label{eq:23}
\Lambda \left( {x,x'} \right) = \left( {\lambda \sqrt {2\pi } } 
\right)^{-1} \cdot \exp \left\{ { - \frac{{\left( {x - x'} \right)^2 }}
{{2\lambda ^2 }}} \right\}
\end{equation}
where $\gamma$ and $\lambda$ describe the characteristics of the measuring devices. Then 
for $\varrho_{out}$ and $J_{out}$ one finds
  \begin{equation}\label{eq:24}
\rho _{out} \,\left( x \right) = \left[ {2\pi \left( {\sigma ^2  + 
\gamma ^2 } \right)} \right]^{ - 1/2}  \cdot \exp \left\{ { - \frac{{\left( 
{x - x_0 } \right)^2 }}{{2\left( {\sigma ^2  + \gamma ^2 } \right)}}} \right\}
\end{equation} 
  \begin{equation}\label{eq:25}
  J_{out} \left( x \right) = \hbar k\left[ {2\pi m^2 \left( {\sigma ^2  + 
  \lambda ^2 } \right)} \right]^{ - 1/2}  \cdot \exp \,\left\{ { - \frac{{\left( 
  {x - x_0 } \right)^2 }}{{2\left( {\sigma ^2  + \lambda ^2 } \right)}}} \right\}
  \end{equation} 
  One can see that in the case when both $\gamma \rightarrow 0$ and $\lambda\rightarrow 0$
  the kernels  $\Gamma (x)$ and $\Lambda (x)$ degenerate into the Dirac's function
  $\delta (x - x')$. Then $\varrho_{out} \rightarrow \varrho_{in}$ and
  $J_{out} \rightarrow J_{in}$. Such a case corresponds to an ideal measurement. 
  Alternatively the cases when $\gamma \neq 0$ and/or $\lambda \neq 0$ are associated with
  non-ideal measurements.
  
  As observables of interest we take the coordinate $x$ and momentum $p$ described by the
  operators $\hat x = x \cdot$ and $\hat p =  - i \hbar \,\frac{\partial }{{\partial x}}$
  Then, according to the scheme presented in the previous section, one obtains
    \begin{equation}\label{eq:26}
	\left\langle x \right\rangle _{in}  = \left\langle x \right\rangle 
	_{out} = x_0 \quad ,\quad \left\langle p \right\rangle _{in} = 
	\left\langle p \right\rangle _{out}  = \hbar  \cdot k
\end{equation} 
  \begin{equation}\label{eq:27}
  \mathcal{C}_{in} \left( {x,p} \right) = \mathcal{C}_{out} \left( {x,p} \right) = \frac{{i\hbar }}{2}
\end{equation}
\begin{equation}\label{eq:28}
\Delta _{in} x = \sigma \quad ,\quad \Delta _{out} x = \sqrt {\sigma ^2 
 + \gamma ^2 } 
\end{equation}
\begin{equation}\label{eq:29}
\Delta _{in} p = \frac{\hbar }{{2\sigma }}
\end{equation}
\begin{equation}\label{eq:30}
\Delta _{out} p = \hbar \sqrt {\frac{{k^2 \left( {\sigma ^2  + 
\gamma ^2 } \right)}}{{\sqrt {\sigma ^4  - \gamma ^4  + 2\gamma ^2 \left( 
{\sigma ^2  + \lambda ^2 } \right)} }} - k^2  + \frac{1}{{4\left( 
{\sigma ^2  + \gamma ^2 } \right)}}} 
\end{equation}
Subsequently for the corresponding uncertainty indicators of predicted (\emph{prd}) type
defined in \eqref{eq:17},\eqref{eq:18} and \eqref{eq:19} one finds
\begin{equation}\label{eq:31}
\varepsilon _{prd} \left( {\left\langle x \right\rangle } \right) = 0 ,\quad 
\varepsilon _{prd}\left( {\left\langle p \right\rangle } \right) = 0,\quad 
\varepsilon _{prd}\left( {\left\langle {\mathcal{C}\left({x,p}\right)} \right\rangle } \right) 
= 0
\end{equation}
\begin{equation}\label{eq:32}
\varepsilon_{prd}\left( {\Delta x} \right) = \sqrt {\sigma ^2  + 
\gamma ^2 } - \sigma 
\end{equation}
\begin{equation}\label{eq:33}
\varepsilon_{prd}\left( {\Delta p} \right) = \Delta_{out} p - 
\Delta_{in} p \ne 0
\end{equation}
These relations show that in the considered model the estimators 
$\left\langle x \right\rangle$, $\left\langle p \right\rangle$ and
$\mathcal{C} \left({x,p}\right)$ have not predicted uncertainties. But within the same model the
estimators $\Delta x $ and $\Delta p$ are characterized by non-null uncertainties.

For an evaluation of the interdependencies between the uncertainties of $x$ and $p$
from \eqref{eq:31}-\eqref{eq:33} one obtains 
\begin{equation}\label{eq:34}
\varepsilon \left( {\left\langle x \right\rangle } \right) \cdot
\varepsilon \left( {\left\langle p \right\rangle } \right) = 0
\end{equation}
\begin{equation}\label{eq:35}
\varepsilon \left( {\Delta x} \right) \cdot \varepsilon \left( 
{\Delta p} \right) = \hbar  \cdot {\rm W}
\end{equation}
Here $W$ is a real, non-negative and dimensionless quantity which can be evaluated
by means of the relations \eqref{eq:33},\eqref{eq:28}, \eqref{eq:29} and \eqref{eq:30}.

If in \eqref{eq:20} we restrict to the values $x = 0$, $k = 0$ and 
$\sigma = \sqrt {\frac{\hbar }{{2m\omega }}}$ our system is just a linear 
oscillator situated in its ground state ($m$ = mass and $\omega$ = pulsation). For an 
observable of interest we refer to the energy $E$ described by the Hamiltonian
\begin{equation}\label{eq:36}
\hat H =  - \frac{{\hbar ^2 }}{{2m}}\frac{{d^2 }}{{dx^2 }} + \frac{{m\omega ^2 }}{2}x^2 
\end{equation}
Then for the probabilistic estimators of energy one finds 
\begin{equation}\label{eq:37}
\left\langle H \right\rangle _{in} = \frac{{\hbar \omega }}
{{2\quad }}\quad ,\quad \Delta_{in} H = 0
\end{equation}
\begin{equation}\label{eq:38}
\left\langle H \right\rangle _{out} = \frac{{\omega \left[ {\hbar ^2 
+ \left( {\hbar + 2m\omega \gamma ^2 } \right)^2 } \right]}}
{{4\left( {\hbar + 2m\omega \gamma ^2 } \right)}}
\end{equation}
\begin{equation}\label{eq:39}
\Delta _{out} H = \frac{{\sqrt 2 m\omega ^2 \gamma ^2 \left( {\hbar + m\omega \gamma ^2 } \right)}}
{{\hbar + 2m\omega \gamma ^2 }}
\end{equation}
The corresponding predicted uncertainty indicators are
\begin{equation}\label{eq:40}
\varepsilon _{prd} \left( {\left\langle H \right\rangle } \right) =
\left\langle H \right\rangle _{out} - \left\langle H \right\rangle _{in} \ne 0
\end{equation}
\begin{equation}\label{eq:41}
\varepsilon_{prd} \left( {\Delta H} \right) = \Delta_{out} H
- \Delta_{in} H \ne 0
\end{equation}
\section{Conclusions}
 
 The problem of QMS description persists in our days as an open and disputed question 
 while many of its approaches are founded on traditional conjectures inspired from TIHR.
 But indubitable facts show \cite{10} that TIUR is an unjustified doctrine. Consequently 
 in this paper we find that the mentioned conjectures imply unsurmountable  deficiencies.
The respective finding motivates our search for a possible new approach of QMS, based on   
reconsidered conjectures. We propose a set of four such conjectures and develop an 
adequate approach of QMS.

Our approach is motivated by some considerations about the characteristics of real QMS. 
So we regard a QMS as a statistical sampling (i.e. as a set of a large number of trials)
which preserves the spectra of the observables but can modify the probabilities associated 
with the values from the respective spectra. Consequently in the proposed approach the 
QM operators of observables are preserved. Concomitantly the probability density and current 
(directly connected with the wave function) are subjected to linear functional 
transformations, from input (in) into output (out) expressions. The respective 
transformations imply the characteristics of measuring devices. The quantum observables
are evaluated through probabilistic estimators such are : mean values, correlations and 
standard deviations. Within the proposed approach the mentioned estimators are 
characterized by both in and out values. Then for each estimator the difference between 
the respective value gives a quantitative evaluation of the measuring uncertainty. A concrete
exemplification of the above ideas is given in section 5.\\
Our approach of QMS is quite different from the approaches founded conjectures inspired 
from TIUR. The difference is evidenced on the one hand by the opinion that a QMS must be 
regarded as a statistical sampling but not as a single trial. Consequently we can avoid the 
controversial idea of wave function collapse (reduction). On the other hand the alluded
difference is pointed out by the presumption that the description of QMS must be regarded 
not as a part of QM theory but as a distinct and independent task comparatively with the 
objectives of usual QM. Note that the opinions promoted here and in \cite{6,7,8,9,10}
in connection with QMS respectively with RSUR are mainly consonant with usual QM. 
Therefore they must not be tested \emph{"against quantum mechanics"} as it is suggested 
\cite{3} for the theories which promote the idea of wave function reduction. 

\section*{Acknowledgments} 
\begin{itemize}
\item  I Wish to express my deep gratitude to those authors, publishing companies 
 and libraries which, during the years, helped me with  copies of some publications
 connected with the problems approached here.
\item The investigations in the field of the present paper benefited partially 
of facilities from the grants supported by the Roumanian Ministry of Education and
 Research.
 \end{itemize}
 
 \section*{List of abbreviations}
\emph{in} = input\\
\emph{fac} = factual\\
\emph{out} = output\\
\emph{prd} = predicted\\
QM = quantum mechanics\\
QMS = quantum measurement(s)\\
\textbf{RC} = reconsidered conjecture\\
RSUR = Robertson Schr\"{o}dinger uncertainty relation\\
\textbf{TC} = traditional conjecture\\
TIUR = traditional interpretation of uncertainty relations\\

\end{document}